\documentclass[12pt]{article}
\pdfoutput=1
\usepackage{putex}
\usepackage{feyn}
\usepackage[vcentermath]{youngtab}
\usepackage{subfig}
\usepackage{lscape}

\usepackage{graphicx}
\usepackage{epstopdf}
\usepackage{enumerate}
\usepackage{cite}
\usepackage{tensor}
\usepackage{slashed}
\usepackage{amsmath}
\usepackage{amssymb}
\usepackage{mathrsfs}
\usepackage{lgrind}

\usepackage{bbm}

\usepackage{multirow}

\usepackage{hyperref}

\numberwithin{equation}{section}

\newcommand {\be} {\begin {equation}}
\newcommand {\ee} {\end {equation}}

\newcommand {\bes} {\begin {equation*}}
\newcommand {\ees} {\end {equation*}}

\def\CH{{\cal H}}

\def\CO{{\cal O}}

\newcommand{\beq}{\begin{equation}}
\newcommand{\eeq}{\end{equation}}

\newcommand{\cO}{{\cal O}}

\newcommand{\SM}{\mathit{SM}}

\def\be{ \begin{equation} }
\def\ee{ \end{equation} }

\begin{document}

\institution{PU}{ Department of Physics, Princeton University, Princeton, NJ 08544, USA}
\institution{PCTS}{ Princeton Center for Theoretical Science, Princeton, NJ 08544, USA}
\institution{PGI}{ Princeton Gravity Initiative, Princeton University, Princeton, NJ 08544, USA}
\institution{Carnegie}{ Department of Physics, Carnegie Mellon University, Pittsburgh, PA 15213, USA}

\title{Ginzburg-Landau Description and Emergent Supersymmetry of 
the $(3,8)$ Minimal Model  
}

\authors{{\normalsize Igor R.~Klebanov,\worksat{\PU, \PCTS} Vladimir Narovlansky,\worksat{\PU} Zimo Sun,\worksat{\PGI} and Grigory Tarnopolsky\worksat{\Carnegie}}}

\abstract{A pair of the 2D non-unitary minimal models $M(2,5)$ is known to be equivalent to a variant of the $M(3,10)$ minimal model. We discuss the RG flow from this model to another non-unitary minimal model, $M(3,8)$. This provides new evidence for its previously proposed Ginzburg-Landau description, which is a $\mathbb{Z} _2$ symmetric theory of two scalar fields with cubic interactions. We also point out that 
$M(3,8)$ is equivalent to the $(2,8)$ superconformal minimal model with the diagonal modular invariant.
Using the 5-loop results for theories of scalar fields with cubic interactions, we exhibit the $6-\epsilon$ expansions of the dimensions of various operators. Their extrapolations are in quite good agreement with the exact results in 2D. We also use them to approximate the scaling dimensions in $d=3,4,5$ for the theories in the $M(3,8)$ universality class.  
 }

\date{}
\maketitle

\tableofcontents

\section{Introduction and summary}

In the landmark paper \cite{Belavin:1984vu}, Belavin, Polyakov and Zamolodchikov defined an infinite class of 2D conformal field theories. They are the conformal minimal models 
$M(p,q)$ with central charges $c(p,q)= 1- 6 (p-q)^2/(pq)$,
where $p$ and $q$ are relatively prime positive integers. Among them a distinguished subset are the unitary minimal models $M(p, p+1)$ \cite{Friedan:1983xq}, where $p=3,4,5, \ldots$. For $p=3$ we find the Ising model, for $p=4$ the tricritical Ising model, etc. 

The critical Ising model is well-known to be described by the massless Euclidean $\phi^4$ field theory. This theory is conformal for $d<4$, and in particular in 2D. More generally, Zamolodchikov \cite{Zamolodchikov:1986db} proposed that the Ginzburg-Landau (GL) description of $M(p, p+1)$ with diagonal modular invariants is given by
\be
\label{ZamGL}
S=\int d^d x \bigg ( \frac{1}{2}\left(\partial_{\mu}\phi\right)^2 + {g\over (2(p-1))!} \phi^{2(p-1)}
\bigg )\ ,
\ee
where we assume that all the coefficients of terms $\phi^{2k}$ with $k<p-1$ are tuned to zero at the multi-critical point.
In $d=2$ the coupling $g$ has dimension of mass-squared, so these models flow to strong coupling. However, they become weakly coupled near the upper critical dimensions $d_c(p)=\frac{2(p-1)}{p-2}$, and one can develop the $d_c-\epsilon$ expansions. For example, for the $\phi^4$ theory ($p=3$) Wilson and Fisher \cite{Wilson:1971dc} developed the $4-\epsilon$ expansion that, when continued to $d=2$ gives a good approximation to the exact Ising scaling dimensions in $d=2$: $\Delta_\phi=1/8$ and $\Delta_{\phi^2}=1$.
Similarly, the tricritical Ising model, $M(4,5)$, is described by the $\phi^6$ theory, which admits a $3-\epsilon$ expansion \cite{Hager:2002uq}. On the other hand, 
the fermionic version \cite{Hsieh:2020uwb} of the tricritical Ising model may be described by a Yukawa theory with emergent Supersymmetry \cite{Fei:2016sgs}. This Yukawa model admits a $4-\epsilon$ expansion, and in 3D it describes the supersymmetric Ising model \cite{Grover:2013rc,Fei:2016sgs,Atanasov:2022bpi}.

Another interesting class of minimal models are $M(2, 2k+1)$ with $k=2, 3, \ldots$. 
The first representative is $M(2,5)$, which corresponds to the Yang-Lee (YL) edge singularity \cite{Cardy:1985yy}. Its GL description is provided by the scalar field theory with interaction $\sim i \phi^3$ \cite{Fisher:1978pf}. Indeed, the theory
\be
\label{FishGL}
S=\int d^d x \bigg ( \frac{1}{2}\left(\partial_{\mu}\phi\right)^2 + {g\over 6} \phi^{3} 
\bigg )\ ,
\ee    
has a weakly coupled IR fixed point in $d=6-\epsilon$ with an imaginary $g$. The resulting $\epsilon$ expansion, which is now known up to order $\epsilon^5$ 
\cite{Kompaniets_2021,Borinsky:2021jdb}, provides a good approximation to the exact value $\Delta_\phi=2 h_{1,2}=-2/5$ in 2D.\footnote{See also \cite{Xu:2022mmw} for a recent discussion of subleading contributions in the YL model related to irrelevant operators.}

The GL descriptions of other non-unitary minimal models pose interesting challenges \cite{amslaurea11308,Zambelli:2016cbw,Anninos:2021eit,Lencses:2022ira,Nakayama:2022svf}.
In this paper we explore another field theory with cubic interactions 
\begin{equation} \label{eq:GL_action}
S = \int d^dx \left( \frac{1}{2} (\partial \phi )^2+\frac{1}{2} (\partial \sigma )^2+\frac{g_1}{2} \sigma \phi ^2+\frac{g_2}{6} \sigma ^3\right) .
\end{equation}
In the concluding paragraphs of \cite{Fei:2014xta}, it was suggested that this model with imaginary $g_1$ and $g_2$ provides a GL description of the minimal model $M(3,8)$, and 
we provide several new arguments in favor of this identification.\footnote{The theory \eqref{eq:GL_action} is the case $N=1$ of the $O(N)$ invariant theories introduced in \cite{Fei:2014yja} as a UV completion of the quartic $O(N)$ invariant theories for dimensions $4<d<6$. In $6-\epsilon$ dimensions it was found that only for $N >1038$ do these theories posses fixed points in $d=6-\epsilon $ dimensions with \emph{real} couplings. The case $N=1$ instead has fixed points at imaginary couplings. Similarly, the cubic field theory where $\phi$ is
replaced by a pair of anti-commuting scalars has an imaginary fixed point; it describes the $OSp(1|2)$ symmetric universality class that is critical in $2<d<6$ \cite{Fei:2015kta,Klebanov:2021sos,Caracciolo:2004hz,2021CMaPh.381.1223B,Narovlansky:2022ijq}.} 
First, as explained in Section \ref{sec:minimal_model_38}, the global symmetry (other than the conformal symmetry) of $M(3,8)$ is $\mathbb{Z} _2$. This is also evident in the GL description where under the $\mathbb{Z} _2$ the fields transform as
\begin{equation}
\label{zeetwo}
\textbf{Z}_2: \qquad \phi \to -\phi ,\qquad \sigma \to \sigma .
\end{equation}
There is also an anti-unitary time reversal symmetry (where $i \to -i$). This is shown for $M(3,8)$ in Section \ref{sec:minimal_model_38}. It is also true for \eqref{eq:GL_action}, where at the fixed point the couplings are imaginary. Under this symmetry the fields transform as
\begin{equation} \label{eq:GL_time_reversal}
T: \qquad \phi \to \phi ,\qquad \sigma \to -\sigma .
\end{equation}
In \cite{Fei:2014xta} it was suggested that, when $M(3,10)$ is perturbed by the maximum dimension relevant operator with $\Delta=6/5$, it can flow to $M(3,8)$. Here we show that this flow is consistent with the generalized $c$-theorem that applies to all 
PT-symmetric models \cite{Castro-Alvaredo:2017udm}. 
The $D_6$ modular invariant model of $M(3,10)$ describes a pair of the Yang-Lee models $M(2,5)$ \cite{Kausch:1996vq,Quella:2006de,2011NJPh...13d5006A}. Indeed, the central charge $c(3,10)= 2 c(2,5)=-44/5$, and the operator scaling dimensions are sums of those found in $M(2,5)$.
We therefore suggest that the GL description of $M(3,10)$ is provided by two copies of the cubic field theory (\ref{FishGL}) with the same couplings. 
This GL description has $\mathbb{Z} _2$ symmetry realized by interchanging the two scalar fields, $\phi_1$ and $\phi_2$. After defining $\phi_1= \frac{\sigma+\phi}{\sqrt 2}$,
$\phi_2= \frac{\sigma-\phi}{\sqrt 2}$, we again find the GL description of (\ref{eq:GL_action}) with $g_1=g_2$. Therefore, the RG flow from $M(3,10)$ to $M(3,8)$ can be studied using this GL description, generated by the operator $\phi _1\phi _2^2+\phi _2\phi _1^2$. 
This $\mathbb{Z} _2$ invariant flow can be studied perturbatively in $6-\epsilon$ dimensions, where it connects two different fixed points \cite{Fei:2014yja}.

In Section \ref{sec:SUSY} we recall that the minimal model $M(3,8)$ has emergent supersymmetry: it is the $(2,8)$ member of the class of ${\cal N}=1$ superconformal minimal models
\cite{Melzer:1994qp,Nakayama:2021zcr}.
This is analogous to the well-known equivalence between the tricritical Ising model $M(4,5)$ and the superconformal $(3,5)$ minimal model with the diagonal modular invariant.
The $\mathbb{Z} _2$ symmetry (\ref{zeetwo}) translates in the superconformal $(2,8)$ model into the symmetry which acts by $(-1)$ on the R-R sector and by $+1$ on the NS-NS sector.  
We note that it seems challenging to find a Ginzburg-Landau description for the fermionic superconformal $(2,8)$ model \cite{Nakayama:2021zcr}, and there are indications that there is no field theory in which we can apply the $\epsilon $-expansion to describe this model.\footnote{The reason is essentially that there is no choice of a monomial superpotential that would be consistent with the symmetries of the model.} Interestingly, here we observe that if we restrict to the bosonic $M(3,8)$ model, there is such a GL description.
In the remainder of the paper, we use the GL description (\ref{eq:GL_action}) and the perturbative results
\cite{Fei:2014xta,Gracey:2015tta,Fei:2015kta,Kompaniets_2021,Borinsky:2021jdb}
to carry out the $6-\epsilon$ expansions of the operator dimensions. We find that, when extrapolated to 2D, they are in reasonably good agreement with the exact dimensions of $M(3,8)$.

\section{The $(3,8)$ and $(3,10)$ minimal models and RG flows} \label{sec:minimal_model_38}

We start by mentioning the basic structure of the $(3,8)$ minimal model that is of interest to us. This is the theory consisting of two, left and right, copies of the corresponding degenerate representation of the Virasoro algebra. We are considering the diagonal theory, where each primary field has the same holomorphic and anti-holomorphic dimensions. The central charge of the theory (left and right) is $c=-\frac{21}{4} $. There are seven primary operators which we will denote by $\phi _{1,i} $, $i=1,\cdots ,7$, where $\phi _{1,1} = 1$ is the identity. Their conformal dimensions are shown in Table \ref{tab38_MM_primaries}.

\begin{table}[h]
\centering
\begin{tabular}{c | c c c c c c} 
$M(3,8)$ & $\phi _{1,2}$ & $\phi _{1,3}$ & $\phi _{1,4}$ & $\phi _{1,5}$ & $\phi _{1,6}$ & $\phi _{1,7}$ \\
 \hline
$\Delta $ & $-\frac{7}{16}\approx -0.44 $ & $-\frac{1}{2} $ & $-\frac{3}{16} \approx -0.19$ & $\frac{1}{2} $ & $ \frac{25}{16} \approx 1.56$ & $3$ \\
$\mathbb{Z} _2$ & odd& even &odd & even & odd & even \\
T & even & odd & odd & odd & even & even
\end{tabular}
\caption{Primary fields and their properties for the $(3,8)$ minimal model. $\Delta $ denotes the conformal dimension, which is the sum of the holomorphic and the anti-holomorphic dimensions.}
\label{tab38_MM_primaries}
\end{table}

The OPE structure of the theory is given by
\begin{equation} \label{eq:MM_OPEs}
\begin{aligned}
& \phi _{1,2} \times \phi _{1,2} \sim 1+i\phi _{1,3} , \qquad && \phi _{1,2} \times \phi _{1,3} \sim i \phi _{1,2} +\phi _{1,4} ,\\
& \phi _{1,2} \times \phi _{1,4} \sim \phi _{1,3} +\phi _{1,5} , \qquad && \phi _{1,2} \times \phi _{1,5} \sim \phi _{1,4} +i\phi _{1,6} ,\\
& \phi _{1,2} \times \phi _{1,6} \sim i \phi _{1,5} +\phi _{1,7} ,\qquad && \phi _{1,2} \times \phi _{1,7} \sim \phi _{1,6} ,\\
& \phi _{1,3} \times \phi _{1,3} \sim 1+i \phi _{1,3} +i \phi _{1,5} ,\qquad && \phi _{1,3} \times \phi _{1,4} \sim \phi _{1,2} +i \phi _{1,4} +\phi _{1,6} ,\\
& \phi _{1,3} \times \phi _{1,5} \sim i \phi _{1,3} +i \phi _{1,5} +\phi _{1,7} ,\qquad && \phi _{1,3} \times \phi _{1,6} \sim \phi _{1,4} +i \phi _{1,6} ,\\
& \phi _{1,3} \times \phi _{1,7} \sim \phi _{1,5} ,\qquad && \phi _{1,4} \times \phi _{1,4} \sim 1+i \phi _{1,3} +i \phi _{1,5} +\phi _{1,7} ,\\
& \phi _{1,4} \times \phi _{1,5} \sim \phi _{1,2} +i \phi _{1,4} +\phi _{1,6} ,\qquad && \phi _{1,4} \times \phi _{1,6} \sim \phi _{1,3} +\phi _{1,5} ,\\
& \phi _{1,4} \times \phi _{1,7} \sim \phi _{1,4} ,\qquad && \phi _{1,5} \times \phi _{1,5} \sim 1+i \phi _{1,3} +i \phi _{1,5} ,\\
& \phi _{1,5} \times \phi _{1,6} \sim i \phi _{1,2} +\phi _{1,4} ,\qquad && \phi _{1,5} \times \phi _{1,7} \sim \phi _{1,3} ,\\
& \phi _{1,6} \times \phi _{1,6} \sim 1+i \phi _{1,3} ,\qquad && \phi _{1,6} \times \phi _{1,7} \sim \phi _{1,2} ,\\
& \phi _{1,7} \times \phi _{1,7} \sim 1 .
\end{aligned}
\end{equation}
Note that there are no degeneracies in the dimensions of the fields. The global (0-form) symmetries are dictated by invariance of the OPE. In this case, the only such symmetry commuting with Virasoro is a $\mathbb{Z} _2$ symmetry, under which $\phi _{1,2} $, $\phi _{1,4} $, and $\phi _{1,6} $ are odd, as indicated in Table \ref{tab38_MM_primaries}.

The OPE coefficients (or 3-point function coefficients) for the minimal model are known as well \cite{Dotsenko:1984nm,Dotsenko:1984ad,Dotsenko:1985hi}. We are particularly interested in
\begin{equation} \label{eq:MM_OPE_coeffs}
\begin{split}
& C_{(1,3),(1,3),(1,3)} ^2 = -\frac{\Gamma \left( \frac{1}{8} \right) ^2 \Gamma \left( \frac{5}{4} \right) }{8 \sqrt{2} \cdot  \Gamma \left( \frac{7}{8} \right) ^2 \Gamma \left( \frac{3}{4} \right) } \approx -3.12537,\\
& C_{(1,2),(1,2),(1,3)} ^2 = -\frac{ \sqrt{2}\cdot \Gamma \left( \frac{1}{8} \right) ^2 \Gamma \left( \frac{5}{4} \right) }{8 \cdot  \Gamma \left( \frac{7}{8} \right) ^2 \Gamma \left( \frac{3}{4} \right) }=2 C_{(1,3),(1,3),(1,3)} ^2  \approx -6.25073 ,
\end{split}
\end{equation}
where for instance the first quantity corresponds to the 3-point function (including both holomorphic and anti-holomorphic pieces) coefficient of three copies of $\phi _{1,3} $.
In particular, both OPE coefficients are imaginary.
We will not quote all coefficients, but we have indicated which couplings are imaginary by writing an `$i$'  in \eqref{eq:MM_OPEs}.

In addition to such global $\mathbb{Z} _2$ symmetries, there can also be spacetime symmetries, and in particular a time reversal symmetry. Indeed, there is such a time reversal symmetry for $M(3,8)$ with the charges indicated in Table \ref{tab38_MM_primaries}.\footnote{There is also the symmetry that is the product of the two. In the GL description \eqref{eq:GL_action} it corresponds to transforming both $\phi $ and $\sigma $ in \eqref{eq:GL_time_reversal}. It is not an independent symmetry of course.}

Let us argue that there is RG flow from $M(3,10)$ to $M(3,8)$.
The minimal model $M(3,10)$ has 9 primary fields, with $\phi _{1,1} =1$ being the identity. Their conformal dimensions are shown in Table \ref{tab310_MM_primaries}. More details on $M(3,10)$ are given in Appendix \ref{sec:310_minimal_model}.

Flows between non-unitary but PT-symmetric theories must satisfy a generalized version of the $c$-theorem
\cite{Castro-Alvaredo:2017udm},
which states that the effective central charge $c_{\rm eff}= c - 24 h_{\rm min}$ must decrease. For the unitary theories, $h_{\rm min}=0$ and we recover the original Zamolodchikov $c$-theorem \cite{Zamolodchikov:1986gt}. For the conformal minimal models $c_{\rm eff}(p,q)=1- 6/(pq)$, and its decrease imposes strong constraints on possible RG flows. For example, for the flows starting from $M(4,5)$ these constraints were recently used in \cite{Lencses:2022ira}. 

\begin{table}[h]
\centering
\begin{tabular}{c | c c c c c c c c} 
$M(3,10)$ & $\phi _{1,2}$ & $\phi _{1,3}$ & $\phi _{1,4}$ & $\phi _{1,5}$ & $\phi _{1,6}$ & $\phi _{1,7}$ & $\phi _{1,8} $ & $\phi _{1,9} $ \\
 \hline
$\Delta $ & $-\frac{11}{20} $ & $-\frac{4}{5} $ & $-\frac{3}{4} $ & $-\frac{2}{5} $ & $ \frac{1}{4} $ & $\frac{6}{5}$ & $\frac{49}{20} $ & $4$ \\
$\mathbb{Z} _2$ & odd& even &odd & even & odd & even & odd & even \\
T & even & even & odd & odd & odd & even & even & even
\end{tabular}
\caption{Primary fields and their properties for the $(3,10)$ minimal model.}
\label{tab310_MM_primaries}
\end{table}

Since  $c_{\rm eff}(3,10)=4/5$, we find that from the flows originating from $M(3,10)$ the IR minimal model must have $pq<30$.
Given that this is a non-unitary minimal model, this limits the possibilities to $(2,5)$, $(2,7)$, $(2,9)$, $(2,11)$, $(2,13)$, $(3,5)$, $(3,7)$, $(3,8)$, $(4,7)$.
Furthermore, since $M(3,10)$ has $7$ primary relevant operators (including the identity), we will assume that the IR theory has $6$ of them. Except for $M(2,13)$ that we will rule out in a moment, the only such minimal model is $M(3,8)$. Since the RG flow takes place within the space of GL theories (\ref{eq:GL_action}) this solidifies the tentative identification of the IR fixed point with $M(3,8)$ that was made in 
\cite{Fei:2014yja}.

In fact, 
all the other possibilities can be ruled out as follows.
The models $(2,5)$, $(2,7)$, $(2,9)$, $(2,11)$, $(2,13)$ have no $\mathbb{Z} _2$ symmetry.
The $(4,7)$ minimal model does have a $\mathbb{Z} _2$ and time reversal symmetry. However, the lowest dimension $\mathbb{Z} _2$ odd operator is $\phi _{2,3} $ and the lowest dimension $\mathbb{Z}_2$ even operator is $\phi _{1,2} $ of the $M(4,7)$ theory. We would therefore expect by \eqref{eq:GL_time_reversal} that $\phi _{1,2} $ of the $M(4,7)$ theory would be odd under time reversal, but in fact it can be checked that it is even. In addition, we would expect the 3-point coefficient of $\phi _{1,2} $ of $M(4,7)$ to be imaginary, but it vanishes. This leaves us only with $M(3,5)$, $M(3,7)$, and $M(3,8)$, and among them the scaling dimensions match best those of $M(3,8)$, as we will see below. In fact, 
$M(3,5)$ and $M(3,7)$ can be excluded because, while they do have a $\mathbb{Z} _2$ symmetry, it is anomalous \cite{Cordova:2019wpi,Nakayama:2022svf}.\footnote{We thank Shu-Heng Shao and the referee for pointing this out to us.} Indeed, for these minimal models there is no $D$-type modular invariant, and the $\mathbb{Z} _2$ cannot be gauged.
This puts our proposal on solid footing.

\subsection{RG flow from a pair of Ising models}
\label{twoIsing}

The minimal model $M(2,5)$ describes the YL critical behavior of the Ising model 
in the imaginary magnetic field \cite{Cardy:1985yy}. This implies the existence of an RG flow from the Ising fixed point $M(3,4)$ to the YL  $M(2,5)$  fixed point under the thermal and imaginary magnetic field deformations \cite{Lencses:2022ira}. 
Indeed, using the GL description of the Ising model in magnetic field and at a temperature above the critical ($m \propto T-T_{c}>0$)
\begin{align}
S_{\textrm{Ising}} = \int d^{2}x \bigg(\frac{1}{2}(\partial_{\mu}\phi)^{2}+ h\phi+\frac{1}{2}m\phi^{2}+ \frac{\lambda}{4!}\phi^{4} \bigg)
\ ,\end{align} 
the YL model can be obtained by shifting $\phi \to \phi_{0} + \phi$, with $\phi_{0}=i\sqrt{\frac{2m}{\lambda}}$ to eliminate the quadratic term. Then neglecting the fourth-order term in the action one finds \cite{Fisher:1978pf}:
\begin{align}
S_{\textrm{YL}} = \int d^{2}x \bigg(\frac{1}{2}(\partial_{\mu}\phi)^{2}+ (h+ih_{c})\phi+ \frac{g}{6}\phi^{3} \bigg)\,,
\end{align} 
where $h_{c}=\frac{2m}{3} \sqrt{\frac{2m}{\lambda}}$ and $g =i\sqrt{2m\lambda} $. Tuning the magnetic field $h$ to the imaginary critical value $-ih_{c}$ we obtain the GL description
(\ref{FishGL}) of the YL critical point.    Similarly there is an RG flow from two decoupled Ising models $M(3,4)$ to two decoupled  YL models $M(2,5)$, or equivalently
$M(3,10)$. 

Let us argue that  there exists an RG flow from a pair of Ising models to the $M(3,8)$ model that is produced by the imaginary magnetic field, thermal, spin-spin and energy-energy deformations. Indeed,  we can consider the GL  description of two Ising models at a temperature above the critical and in imaginary field and with spin-spin and energy-energy deformations:
\begin{align}
S =\int d^{2}x \bigg(\sum_{i=1}^{2}\Big(\frac{1}{2}(\partial_{\mu}\phi_{i})^{2}+ h \phi_{i}+\frac{m}{2}\phi^{2}_{i}+ \frac{\lambda}{4!}\phi_{i}^{4}\Big) + \alpha \phi_{1}\phi_{2}+\frac{u}{4} \phi_{1}^{2}\phi_{2}^{2}  \bigg)\,.
\end{align} 
To eliminate the quadratic terms we make a pure imaginary shift of the order parameters: $\phi_{1} \to \phi_{0} + \phi_{1}$ and $\phi_{2} \to \phi_{0} + \phi_{2}$, where $\phi_{0}=i \sqrt{\frac{2m}{\lambda+u}}$. Then taking $\alpha = \frac{2mu}{\lambda +u}$ we obtain
\begin{align}
S =\int d^{2}x \bigg(\sum_{i=1}^{2}\Big(\frac{1}{2}(\partial_{\mu}\phi_{i})^{2}+ (h+ih_{c})\phi_{i} +\frac{g}{6}\phi_{i}^{3}\Big)+  \frac{gu}{2\lambda}\big(\phi_{1}\phi_{2}^{2}+\phi_{2}\phi_{1}^{2}\big)
\bigg)\,,
\end{align} 
where $h_{c}=\frac{2(\lambda+3u)}{3(\lambda+u)}m \sqrt{\frac{2m}{\lambda+u}}$ and $g= i\sqrt{\frac{2m \lambda^{2}}{\lambda+u}}$. At the critical value $h = -ih_{c}$ of the imaginary magnetic filed and after the replacement $\phi_{1}=\frac{\sigma+\phi}{\sqrt{2}}$ and $\phi_{2}=\frac{\sigma-\phi}{\sqrt{2}}$ we find the proposed GL description (\ref{eq:GL_action}) of the minimal model $M(3,8)$,
where $g_{1}=g \frac{\lambda-u}{\sqrt{2}\lambda}$ and $g_{2}= g \frac{\lambda+3u}{\sqrt{2}\lambda}$. As a consistency check, we note that the RG flow from the two Ising models to $M(3,8)$ satisfies the 
inequality \cite{Castro-Alvaredo:2017udm} for $c_{\rm eff}$, since $2 c_{\rm eff}(3,4)=1 > c_{\rm eff}(3,8)=3/4$. This picture suggests that the $M(3,8)$ critical point can be realized by the 2D lattice Hamiltonian of the form 
\begin{align}
H = -\sum_{\langle ij \rangle}(J(\sigma_{1}^{i}\sigma_{1}^{j} +\sigma_{2}^{i}\sigma_{2}^{j} ) + J_{\varepsilon}\sigma_{1}^{i}\sigma_{1}^{j} \sigma_{2}^{i}\sigma_{2}^{j} ) - \sum_{i}  (J_{\sigma} \sigma_{1}^{i}\sigma_{2}^{i} + h (\sigma_{1}^{i}+\sigma_{2}^{i}))\,,
\end{align} 
where $\sigma_{1}^{i},\sigma_{2}^{i} = \pm 1$ are classical spins and the magnetic field $h$ is tuned to critical imaginary value.

\begin{figure}[h!]
\begin{center}
\includegraphics[width=0.75\textwidth]{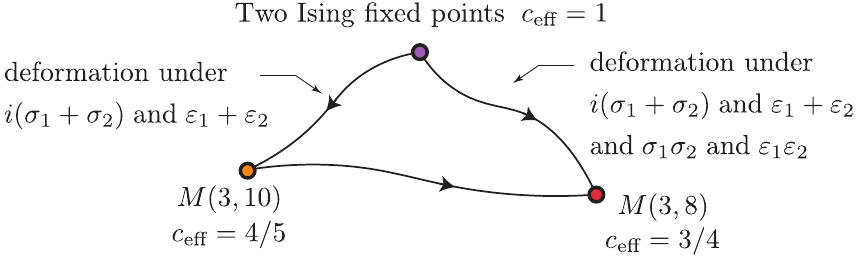}
\end{center}
\caption{A schematic description of the proposed RG flows.}
\label{rgflows}
\end{figure}

The schematic RG flows we are proposing are depicted in Fig.\ \ref{rgflows} (the spin-spin and energy-energy deformations of a pair of Ising models were discussed in \cite{LeClair:1997gv, Delfino:1997ya, Calabrese:2001bm}).

\section{Supersymmetry of $M(3,8)$}
\label{sec:SUSY}

The spectrum of the minimal model $M(3,8)$ suggests that it is a supersymmetric theory. In particular, the holomorphic dimension $3/2$ suggests that the corresponding operator is the supercurrent. In fact, $M(3,8)$ is known to be related to the $(2,8)$ ${\cal N}=1$ supersymmetric minimal model \cite{Melzer:1994qp,Nakayama:2021zcr}. This model is the simplest non-unitary superconformal model; it is sometimes called the supersymmetric Yang-Lee model.\footnote{
We thank the referee for providing a number of references about $\SM(2,8)$ and its relation with $M(3,8)$, and pointing out that this relation was known before.}
Integrable perturbations of it were studied, e.g., in \cite{Schoutens:1990vb,Ahn:1990uq,Ahn:1993qa,Moriconi:1995aj,Ahn:2000tj}

We start by reviewing the ${\cal N}=1$ superconformal minimal models \cite{Eichenherr:1985cx,Bershadsky:1985dq,Friedan:1984rv,Cappelli:1986ed,DiFrancesco:1988xz} (see also \cite{Klebanov:2003wg} for a concise summary).
These are labeled by pairs $(p,q)$, where we can restrict to $p<q$, such that $p,q$ have the same parity with $p$ and $\frac{p-q}{2} $ being coprime. The superconformal central charge is
\begin{equation}
\hat c = 1-\frac{2(p-q)^2}{pq} 
\end{equation}
which is related to the conformal central charge by
$c = \frac{3}{2} \hat c$.
In the superconformal minimal models, $c< \frac{3}{2} $. The primary fields are labeled by $(r,s)$ in the range
\begin{equation}
1 \le r \le p-1,\qquad 1 \le s \le q-1,\qquad sp \le rq.
\end{equation}
The holomorphic dimensions are
\begin{equation}
h_{rs} =\frac{(rq-sp)^2-(p-q)^2}{8pq} +\frac{1-(-1)^{r-s} }{32}.
\end{equation}
The fields with $r-s$ even belong to the Neveu-Schwarz (NS) sector while those with $r-s$ odd are in the Ramond (R) sector. The models with $q=p+2$ are the unitary series. The most familiar one is $\SM(3,5)$, where $\SM(p,q)$ denotes the superconformal minimal models with the diagonal modular invariant. It is equivalent to the tricritical Ising model $M(4,5)$, whose central charge is $c=\frac{7}{10} $.

Let us recall \cite{Melzer:1994qp,Nakayama:2021zcr} that $M(3,8)$ is equivalent to the model $\SM(2,8)$, having $c=-\frac{21}{4} $ or $\hat c=-\frac{7}{2} $. The spectrum of $\SM(2,8)$ is shown in Table \ref{tab28_SMM_primaries}.\footnote{The model $\SM(2,8)$ coupled to 2D supergravity, and its dual large $N$ matrix model description, were discussed in \cite{Klebanov:2003wg}.}
In order to distinguish fields of different minimal models, we will add superscripts to indicate the theory. Let us now match the different fields. As usual, the identity operator in $\SM(2,8)$ gives rise to the identity in $M(3,8)$ as well as to the supercurrent $\phi ^{M(3,8)} _{1,7} $ which is a descendant in $\SM(2,8)$. All other NS fields give rise to two Virasoro primaries. Indeed, $\phi _{1,3} ^{\SM(2,8)} $ gives rise to $\phi _{1,3} ^{M(3,8)} $ having the same dimension, as well as $\phi _{1,5} ^{M(3,8)} $ which is an NS descendant with additional $(\frac{1}{2} ,\frac{1}{2} )$ holomorphic dimensions.

\begin{table}[h]
\centering
\begin{tabular}{c | c c c c c} 
$\SM(2,8)$  & $\phi _{1,1} =1$ & $\phi _{1,2}$ & $\phi _{1,3}$ & $\phi _{1,4}$ \\
 \hline
$\Delta $ & 0 & $-\frac{3}{16} $ & $-\frac{1}{2} $ & $-\frac{7}{16} $ \\
Sector & NS-NS & R-R & NS-NS & R-R
\end{tabular}
\caption{Primary fields and their properties for the $(2,8)$ superconformal minimal model. $\Delta $ denotes the conformal dimension.}
\label{tab28_SMM_primaries}
\end{table}

Usually, each R sector superconformal primary gives rise to a single Virasoro primary. Indeed, $\phi _{1,2} ^{\SM(2,8)} $ corresponds to $\phi _{1,4} ^{M(3,8)} $ having the same dimension, and $\phi _{1,4} ^{\SM(2,8)} $ has the same dimension as $\phi _{1,2} ^{M(3,8)} $. However, there is an exception to this rule. In order to explain this, recall that the generators of the superconformal algebra are $L_n$, the Virasoro generators, and $G_m$, where $m$ are integer in the R sector and half-integer in the NS sector. They satisfy the algebra
\begin{equation}
\begin{split}
& [L_m,L_n]=(m-n)L_{m+n} +\frac{\hat c}{8} m(m^2-1)\delta _{m,-n} ,\\
& \{G_m,G_n\}=2L_{m+n} +\frac{\hat c}{8} \left( 4m^2-1 \right) \delta _{m,-n} ,\\
& [L_m,G_n]=\left( \frac{m}{2} -n\right) G_{m+n} .
\end{split}
\end{equation}
In the R sector we have $G_0^2 = L_0-\frac{\hat c}{16} $. For the R sector superconformal primaries $\phi$, we find that in the very special case that the holomorphic dimension is $h =\frac{c}{24} =\frac{\hat c}{16} $, the operator $G_{-1} \phi $ is a Virasoro primary rather than a descendant. This is the case for the model $\SM(2,8)$, since the operator $\phi _{1,4} ^{\SM(2,8)} $ satisfies this condition, as it has $h =-\frac{7}{32} $. In fact, this applies to all $\SM(p,q)$ where $p$ and $q$ are even, since the operator 
$\phi _{p/2,q/2}$ satisfies this condition.\footnote{The models $\SM(p,q)$ where $p$ and $q$ are even therefore have a non-vanishing Witten index; this means that they have unbroken supersymmetry \cite{Witten:1982df}. In contrast, the models where $p$ and $q$ are
odd do not contain a superconformal primary with $h=\frac{\hat c}{16} $; they have a vanishing Witten index and the supersymmetry is spontaneously broken.}
In $\SM(2,8)$ this results in a superconformal descendant which is a Virasoro primary, having dimension $h=-\frac{7}{32} +1=\frac{25}{32} $, which corresponds to $\phi _{1,6} ^{M(3,8)} $. This completes the correspondence between the primary fields in the two theories. The operators in the R sector are the ones odd under the $\mathbb{Z} _2$ that we mentioned in Table \ref{tab38_MM_primaries}. 

We note that, in the diagonal modular invariant of $\SM(2,8)$, we have a truncation of the full supersymmetric theory that contains no fields of half-integer spin. In fact this combination is identical to the modular invariant of the $M(3,8)$ minimal model. To show this, let us discuss the superconformal characters  (additional details may be found in Appendix \ref{charproof}). For each NS sector superconformal primary field there are $2$ characters: 
\begin{equation}
\chi^{NS}(\mathfrak q) = \tr  {\mathfrak q}^{L_0-\frac{c}{24}}\ , \qquad 
\widetilde \chi^{NS}(\mathfrak q) = \tr  (-1)^F {\mathfrak q}^{L_0-\frac{c}{24}}\ ,
\end{equation}
where the trace is over all the superconformal descendants.
In the model $\SM(2,8)$, we have the following relations to the Virasoro characters $\chi_{r,s}^V(\mathfrak q)\equiv \tr_{r, s} \mathfrak q^{L_0-\frac{c}{24}}$ in
$M(3,8)$:
\begin{equation}
\begin{split}
& \chi_{1,1}^{NS}(\mathfrak q)= \chi_{1,1}^V(\mathfrak q) + \chi_{1,7}^V(\mathfrak q)\ , \qquad 
\widetilde \chi_{1,1}^{NS}(\mathfrak q)= \chi_{1,1}^V(\mathfrak q) - \chi_{1,7}^V(\mathfrak q)\ , \\ 
& \chi_{1,3}^{NS}(\mathfrak q)= \chi_{1,3}^V(\mathfrak q) + \chi_{1,5}^V(\mathfrak q)\ , \qquad 
\widetilde \chi_{1,3}^{NS}(\mathfrak q)= \chi_{1,3}^V(\mathfrak q) - \chi_{1,5}^V(\mathfrak q)\ . \\
\end{split}
\end{equation}
In the R sector we find
\begin{equation}
\begin{split}
& \chi_{1,2}^{R}(\mathfrak q)= \sqrt 2 \chi_{1,4}^V (\mathfrak q) \ , \qquad \qquad \qquad \widetilde \chi_{1,2}^{R}(\mathfrak q)= 0\ , \\
& \chi_{1,4}^{R}(\mathfrak q)= \chi_{1,2}^V(\mathfrak q) + \chi_{1,6}^V(\mathfrak q)\ , \qquad \quad
\widetilde \chi_{1,4}^{R}(\mathfrak q)= \chi_{1,2}^V(\mathfrak q) - \chi_{1,6}^V(\mathfrak q)= 1\ .\\ 
\end{split}
\end{equation}
Following \cite{Cappelli:1986ed}, the factor of $\sqrt 2$ is included in $\chi_{1,2}^{R}(\mathfrak q)$ because all the contributing states are doubly degenerate.
The simplicity of the characters $\widetilde \chi_{1,2}^{R}$ and $\widetilde \chi_{1,4}^{R}$ is due to the unbroken supersymmetry of $\SM(2,8)$, and the fact that the superconformal primary $\phi_{1,4}^{SM(2,8)}$ corresponds to the unique R ground state which contributes to the Witten index. The diagonal modular invariant of $\SM(2,8)$ is 
\begin{equation}
Z_{\SM(2,8)}= \frac{1}{2} \left ( |\chi_{1,1}^{NS}(\mathfrak q)|^2 +  |\widetilde \chi_{1,1}^{NS}(\mathfrak q)|^2 + |\chi_{1,3}^{NS}(\mathfrak q)|^2 +
|\widetilde \chi_{1,3}^{NS}(\mathfrak q)|^2 +  |\chi_{1,2}^{R}(\mathfrak q)|^2 
+ |\chi_{1,4}^{R}(\mathfrak q)|^2 + 1 \right )\ , \\ 
\end{equation}
and it indeed equals the modular invariant of $M(3,8)$:
\begin{equation}
Z_{M(3,8)}=  \sum_{m=1}^7 |\chi_{1,m}^{V} (\mathfrak q)|^2 \ .
\end{equation}
This provides an explicit check of our identification between the two models.

\section{Pad\'e approximants using the $5$-loop results} \label{sec:5_loop}

In this section we match scaling dimensions of the Ginzburg-Landau theory \eqref{eq:GL_action} and $M(3,8)$ finding a nice agreement.

The cubic theory (\ref{eq:GL_action}) has critical dimension $d_{c}=6$. For $d=6-\epsilon<6$, the theory is renormalizable and its renormalization up to five loops has been carried out by \cite{Kompaniets_2021}. In particular, \cite{Kompaniets_2021} computed the beta functions $\{\beta_{1}, \beta_{2}\}$ of the two couplings $g_1$ and $g_2$, the anomalous dimensions $\{\gamma_\phi, \gamma_\sigma\}$ of the two fundamental fields $\phi$ and $\sigma$, and the operator mixing matrix $\gamma^{(2)}_{ij}$ of the two $\mathbb Z_2$ even quadratic operators $\phi^2$ and $\sigma^2$. The $\mathbb Z_2$ odd quadratic operator $\sigma\phi$ is an exact descendant of $\phi$ at fixed points due to the equation of motion $\partial^2\phi=g_1\sigma\phi$.

Using the functions $\beta_1$ and $\beta_2$, we can first find the fixed point of interest. More explicitly, let us define
\begin{align}
g_1=\sqrt{6\epsilon(4\pi)^3}\, x(\epsilon),\,\,\,\,\, g_2=\sqrt{6\epsilon(4\pi)^3}\,y(\epsilon),
\end{align}
where $x(\epsilon)=\sum_{a=0}^4 x_a \epsilon^a+\CO(\epsilon^5)$ and $y(\epsilon)=\sum_{a=0}^4 y_a \epsilon^a+\CO(\epsilon^5)$ are $\epsilon$ expansions.
 At leading order, the relation $\beta_1=\beta_2=0$ yields two coupled equations for $x_0$ and $y_0$
\begin{align}\label{leadingbeta}
-7x_0^2-12 x_0\, y_0 +y_0^2=1, \,\,\,\,\,  12 x_0^3-3 x_0^2\, y_0+9 y_0^3+y_0=0
\end{align}
The solutions of eq.\ (\ref{leadingbeta}) that correspond to a stable fixed point are \cite{Fei:2014xta}
\begin{align}
y_0^*=\frac{6}{5}x_0^*=\pm\frac{6\, i}{\sqrt{499}}
\end{align}
where the two signs are equivalent and we still stick to the ``+'' sign for simplicity. Once $x_0^*$ and $y_0^*$ are fixed, the rest $x_a, y_a$ can be solved recursively by using  higher order terms in $\beta_1=\beta_2=0$. Altogether, we find 
\small
\begin{align}\label{5fixedpt}
&g_1^*\approx i\sqrt{6\epsilon(4\pi)^3}\left(0.223831+0.0788991 \epsilon-0.0395812 \epsilon ^2+0.0701557 \epsilon ^3 -0.181827 \epsilon ^4+\CO(\epsilon^5)\right)\nonumber\\
&g_2^*\approx i\sqrt{6\epsilon(4\pi)^3} \left(0.268597+0.12292 \epsilon-0.0471511 \epsilon ^2+0.104183 \epsilon ^3-0.274325 \epsilon ^4 +\CO(\epsilon^5)\right)
\end{align}
\normalsize

Note that the coupling constants are imaginary. This agrees with \eqref{eq:MM_OPE_coeffs}. We cannot do a quantitative precise comparison, but we see that qualitatively this is consistent with our proposed correspondence.

Plugging these fixed points into the anomalous dimensions $\gamma_\phi, \gamma_\sigma$ yields the $\epsilon$-expansion of the scaling dimensions of $\phi$ and $\sigma$ at this stable fixed point up to $\epsilon^5$:
\small
\begin{align} \label{5phisigma}
\Delta_\phi&=2-0.5501 \epsilon -0.0234476\epsilon ^2+0.0200648 \epsilon ^3-0.0341125 \epsilon ^4+0.0894357\epsilon^5+\CO\left(\epsilon ^6\right)\nonumber\\
\Delta_\sigma&=2-0.561122 \epsilon -0.0358843 \epsilon ^2+0.0236057 \epsilon ^3-0.0451066 \epsilon ^4+0.119965\epsilon^5+\CO\left(\epsilon ^6\right)
\end{align}
\normalsize
By evaluating the eigenvalues  $\lambda_\pm^{(3)}$ of the matrix $\frac{\partial\beta_i}{\partial g_j}$ at $g_1^*$ and $g_2^*$, we obtain the dimensions of two relevant operators in $6-\epsilon$ dimension, arising from the mixing of the two $\mathbb Z_2$ even cubic operators $\phi^2\sigma$ and $\sigma^3$
\small
\begin{align}\label{Dpm}
\Delta_+^{(3)}&=d+\lambda^{(3)}_+=6-0.77319 \epsilon ^2+1.59707 \epsilon ^3 -4.5542 \epsilon ^4 +15.5329 \epsilon ^5+\CO(\epsilon^6)\\
\Delta_-^{(3)}&=d+\lambda^{(3)}_-=6-0.88978 \epsilon+0.0437751 \epsilon ^2 -0.0395877 \epsilon ^3+0.0750536 \epsilon ^4-0.188671 \epsilon ^5+\CO(\epsilon^6)\nonumber
\end{align}
\normalsize
All  terms in eq.\ (\ref{5phisigma}) and eq.\ (\ref{Dpm}) up to order $\epsilon^3$ agree with the three-loop analysis of \cite{Fei:2014xta}.

Using the mixing matrix $\gamma^{(2)}_{ij}$ between $\phi^2$ and $\sigma^2$, we get 
\small
\begin{align}
\Delta^{(2)}_-&=4-1.0501 \epsilon+0.0116011 \epsilon ^2 +0.0195235 \epsilon ^3-0.0210404 \epsilon ^4+0.0430057 \epsilon ^5+\CO(\epsilon^6)
\end{align}
\normalsize
and $\Delta^{(2)}_+=2+\Delta_\sigma$, which is consistent with the equation of motion $\partial^2\sigma=\frac{g_1}{2}\phi^2+\frac{g_2}{2}\sigma^2$. So $\Delta^{(2)}_+$ corresponds to a scalar operator $\CO_+^{(2)}$ which is a descendant of $\sigma$ at level 2 in any $d=6-\epsilon$, i.e., $\CO_+^{(2)}\sim P^2 \sigma$. The order $\epsilon$ and $\epsilon^2$ terms in $\Delta^{(2)}_\pm$ can also be extracted from  \cite{Fei:2014yja,Fei:2015kta}.

To compare with the conformal data of $M(3,8)$, we will use Pad\'e approximants to extrapolate the $\epsilon$-expansions of scaling dimensions above to $\epsilon=4$, i.e., $d=2$. Namely, given any perturbative series $f(\epsilon)=f_0 +f_1\epsilon+\cdots + f_k \epsilon^{k}+\CO(\epsilon^{k+1})$, we approximate it by a rational function 
\begin{align}
\text{Pade}_{[m,n]}(\epsilon)=\frac{a_0+a_1\epsilon+\cdots +a_m \epsilon^m}{1+b_1\epsilon+\cdots+b_n\epsilon^n}
\end{align}
with $m+n=k$, such that (i) $\text{Pade}_{[m,n]}(\epsilon)$ does not have a pole along $0<\epsilon<4$ and (ii) its small $\epsilon$ expansion matches $f(\epsilon)$ up to $\epsilon^k$.
Then we use $\text{Pade}_{[m,n]}(4)$ as the value of the asymptotic series $f(\epsilon)$ at $\epsilon=4$. We compute both 4-loop Pad\'e, which means truncating the $\epsilon$ expansions up to $\epsilon^4$, and 5-loop Pad\'e. The 4-loop and 5-loop Pad\'e results are based on $\text{Pade}_{[2,2]}$ and  $\text{Pade}_{[3,2]}$ respectively if there are no poles between $0$ and $4$. For $\Delta^{(+)}_3$, we find that $\text{Pade}_{[2,2]}$ has a pole around $\epsilon\approx 1.64632$ and hence $\text{Pade}_{[1,3]}$ will be used. Similarly for $\Delta_-^{(2)}$,  because $\text{Pade}_{[3,2]}$ has a pole around $\epsilon\approx 2.84547$, we provide its $\text{Pade}_{[2,3]}$ approximant.  Altogether, the Pad\'e results are summarized in Table \ref{xx}. 

\begin{table}[h]
\centering
\begin{tabular}{c c | c c c c c } 
 && $\phi $ & $\sigma$ & $\CO^{(2)}_-$ & $\CO^{(3)}_+$ & $\CO^{(3)}_-$  \\
 \hline
 \multirow{2}{*}{$\Delta$} & 4-loop & $-0.423502 $ & $-0.69918 $& 0.187805  & 3.49691 & 2.81839  \\
                                        & 5-loop & $-0.309357 $ & $-0.485249 $& 0.479018 & $4.26805$ & $2.66414$   \\
$\mathbb{Z} _2$ &charge   & odd & even & even & even & even
\end{tabular}
\caption{Pad\'e approximants for the scaling dimensions of $\phi, \sigma,\CO^{(2)}_-, \CO^{(3)}_\pm$ in $d=2$. $\CO^{(3)}_\pm$ arise from the mixing between $ \phi^2\sigma$ and  $\sigma^3$, and $\CO^{(2)}_-$ is a combination of $\phi^2$ and $\sigma^2$.}
\label{xx}
\end{table}

Combining the numerical results of scaling dimensions and the $\mathbb Z_2$ charges, we propose to make the following identifications of primary operators in $M(3,8)$:
\begin{align}
\phi_{1,2}\leftrightarrow \phi, \,\,\,\,\,\phi_{1,3}\leftrightarrow \sigma, \,\,\,\,\, &\phi_{1,5}\leftrightarrow \CO^{(2)}_-,\,\,\,\,\, \phi_{1,7}\leftrightarrow \CO^{(3)}_-
\end{align}
In addition, we think that $\CO_+^{(3)}$ can become a Virasoro descendant of $\sigma$, i.e., $\CO_+^{(3)}\sim L_{-2}\bar L_{-2}\sigma$.

The two $\mathbb Z_2$ odd primaries $\phi_{1,4}$ and $\phi_{1,6}$ are still missing in this dictionary. The first guess would be the two $\mathbb Z_2$ odd cubic operators $\phi^3$ and $\phi\sigma^2$. A simple way to obtain their operator mixing is considering a generalization of (\ref{eq:GL_action}) defined by the potential 
\begin{align}
V(\phi,\sigma)=\frac{g_1}{2} \sigma \phi ^2+\frac{g_2}{6} \sigma ^3+\frac{g_3}{2}\phi\sigma^2+\frac{g_4}{6}\phi^3
\end{align}
The renormalization of this generalized theory has been computed up to three loops in \cite{Osborn_2018}. Diagonalizing the $4\times 4$ stability matrix $\frac{\partial \beta_i}{\partial g_j}$ at the fixed point $(g_1^*, g_2^*, 0,0)$, we recover $\lambda^{(3)}_\pm$ up to $\epsilon^3$. This is because the generalized theory reduces to (\ref{eq:GL_action}) when the couplings $g_3$ and $g_4$ vanish. We also obtain an exactly vanishing eigenvalue which is due to $\text{SO}(2)$ rotations in the $(\phi,\sigma)$ plane  \cite{Osborn_2018}. The last eigenvalue yields a slightly irrelevant operator in $d=6-\epsilon$ with scaling dimension 
\begin{align}
\Delta=6+0.00300601\epsilon-0.770885\epsilon^2+1.59377\epsilon^3+\CO(\epsilon^4)
\end{align}
The $[2,1]$ Pad\'e gives $\Delta\approx 4.68145$ at $d=2$. So this operator is not very likely to correspond to any primary operators in $M(3,8)$. We have to consider higher order operators that are odd under $\mathbb Z_2$.

\section{$\mathbb{Z} _2$ odd quartic operators}

\begin{figure}[h]
\centering
\includegraphics[width=0.9\textwidth]{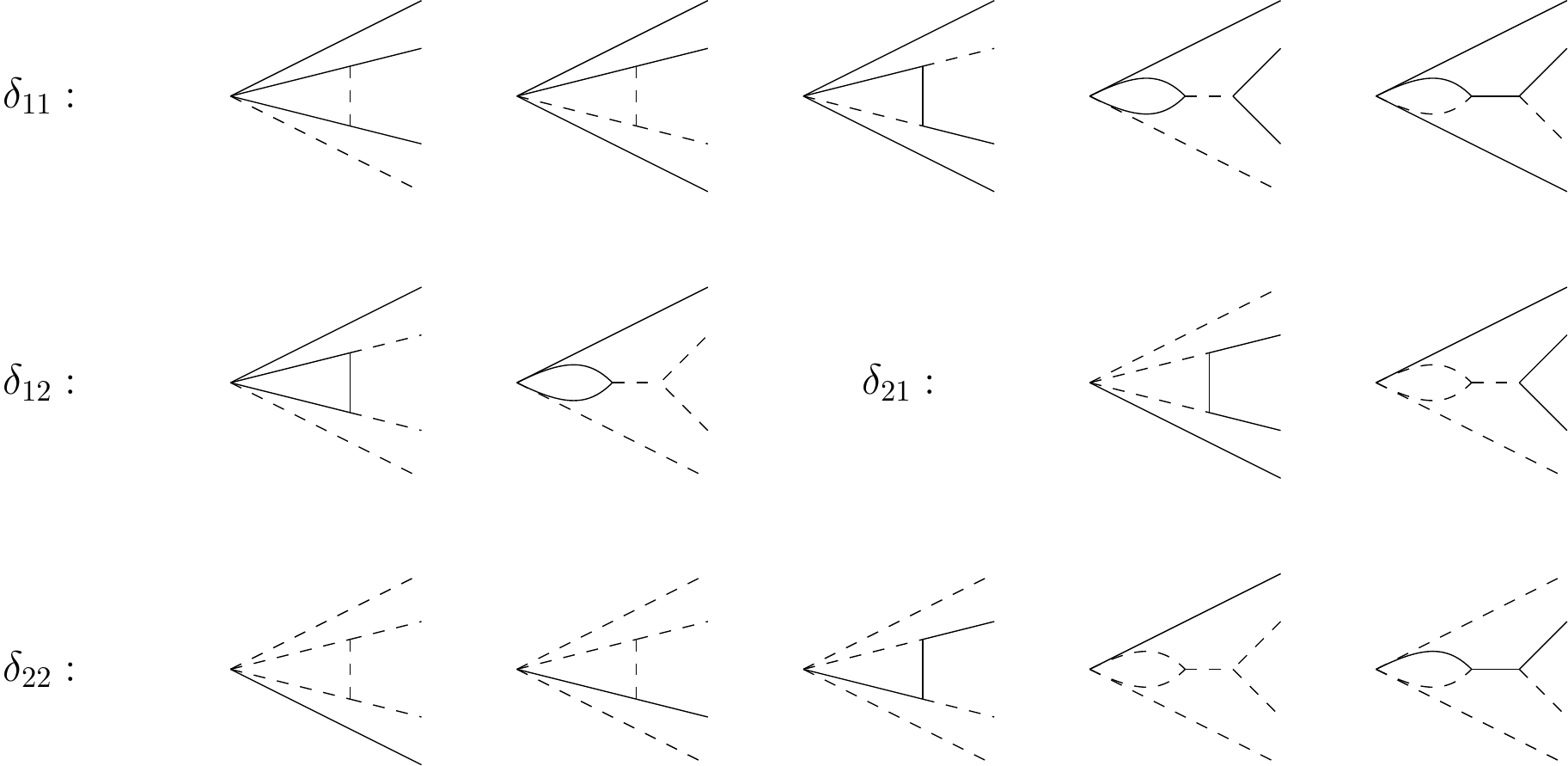}
\caption{Diagrams needed for the calculation of the $\mathbb{Z} _2$ odd quartic operators. We do not show permutations of the same diagrams that are needed.}
\label{fig:quartic_odd_diagrams}
\end{figure}

Next we consider higher $\mathbb{Z} _2$ odd operators, which are the quartic operators $ \frac{\phi ^3 \sigma }{6} $ and $\frac{\phi \sigma ^3}{6} $. We should be careful in doing the calculation in a consistent way from the RG point of view. At the end, the anomalous dimension matrix relevant to obtain the dimensions of these operators is a $2 \times 2$ matrix. One may obtain it by considering the set of diagrams shown in Fig.\ \ref{fig:quartic_odd_diagrams}.

Specifically, one chooses $\delta _{ij} $ with $i,j=1,2$ so as to renormalize the diagrams following each $\delta _{ij} $ shown in the figure. Note that for any permutations of the same diagram, we draw only one of the diagrams. As usual, we should also take into account the anomalous dimensions of the fundamental fields. This means that if we define $\delta _{ij} $ as what we subtract from the diagrams shown, we need to consider the matrix
\begin{equation}
\begin{pmatrix}
1+\delta _{11} +\frac{3}{2} \delta _{\phi } +\frac{1}{2} \delta _{\sigma } & \delta _{12} \\
\delta _{21} & \delta _{22} + \frac{1}{2} \delta _{\phi } +\frac{3}{2} \delta _{\sigma } 
\end{pmatrix} .
\end{equation}
At this order, the wavefunction counterterms are, for example in minimal subtraction,
\begin{equation}
\begin{split}
& \delta  _{\phi } = -\frac{g_1^2}{3(4\pi )^3 \epsilon } ,\\
& \delta _{\sigma } = - \frac{g_1^2+g_2^2}{6(4\pi )^3 \epsilon } .
\end{split}
\end{equation}

The anomalous dimensions matrix is then given by the RG derivative of this matrix as usual. In our case, this gives
\begin{equation}
\gamma^{(4)}  = \frac{1}{(4\pi )^3} 
\begin{pmatrix}
- \frac{47}{12} g_1^2-3g_1g_2+\frac{g_2^2}{12} & -3g_1^2+\frac{g_1g_2}{2} \\
-3g_1^2 + \frac{g_1g_2}{2}  & -\frac{19}{12} g_1^2-3g_1g_2 - \frac{9}{4} g_2^2
\end{pmatrix} .
\end{equation}

The dimensions are obtained by diagonalizing this matrix, giving eigenvalues $\lambda^{(4)} _\pm$ and then as usual
\begin{equation}
\Delta _\pm^{(4)}= 2(d-2)+\lambda ^{(4)}_\pm
\end{equation}
with $d=6-\epsilon $.

Using the fixed point value, we find the dimensions
\begin{equation}
\begin{split}
& \Delta^{(4)} _-=8-\frac{2 \sqrt{135529}-377}{998} \epsilon +O(\epsilon ^2) \approx 8-0.36001\epsilon  , \\
& \Delta^{(4)} _+=8+\frac{2 \sqrt{135529}+377}{998} \epsilon +O(\epsilon ^2) \approx 8+1.11552\epsilon .
\end{split}
\end{equation}

This leading order result is the first step in matching to the minimal model. It is hard to extrapolate as far as $\epsilon =4$ with a 1-loop computation. We hope that higher orders will be computed and matched in the future.

\section{Discussion} 

We have presented new arguments for the existence of RG flow connecting two non-unitary conformal minimal models. The flow connects $M(3,10)$, which is a product of two YL models, in the UV, with $M(3,8)$ in the IR. We have noted that the latter model has emergent supersymmetry: it is equivalent to the superconformal minimal model $\SM(2,8)$.

The GL description (\ref{eq:GL_action}) of $M(3,8)$ and $M(3,10)$ implies that the upper critical dimensions of these minimal models is $6$. The RG flow connecting them can be studied perturbatively in $6-\epsilon$ dimensions, but it becomes strongly coupled in 2D. 
In the future it would be interesting to study the 2D non-unitary RG flow from $M(3,10)$ to $M(3,8)$ numerically using the Truncated Conformal Space Approach \cite{Yurov:1989yu},
which in
\cite{Xu:2022mmw,Lencses:2022ira} was recently applied to other non-unitary RG flows. Similarly, using the methods of \cite{PhysRevB.102.014426}, we can study the RG flow to $M(3,8)$ starting from a pair of Ising models, which was described in section \ref{twoIsing}.

One application of the GL description (\ref{eq:GL_action}) is that it allows us to study the continuation of the $(3,8)$ minimal model to dimension $d>2$
and study the non-trivial CFTs in $d=3,4$, and $5$.\footnote{This is different from the continuation of $\SM(2,8)$ to $d>2$ that was studied in 
\cite{Nakayama:2021zcr}.}
 For each operator, we use the same Pad\'e as in Sec.\ \ref{sec:5_loop}, and present the results for both the 4-loop and the 5-loop computations in Table \ref{tab38_MM_primaries_higher_d}. If the 6-loop results become available, they should provide further improvements to these estimates.

\begin{table}[h]
\centering
\begin{tabular}{c c | c c c c c c} 
&  & $\phi  \leftrightarrow \phi _{1,2}$ & $\sigma \leftrightarrow \phi _{1,3}$ & $\cO ^{(2)} _- \leftrightarrow  \phi _{1,5}$ & $ \cO ^{(3)} _- \leftrightarrow \phi _{1,7}$ & $ \cO ^{(3)} _+$ \\
 \hline
\multirow{2}{*}{$d=3$} & 4-loop & 0.222711 & 0.0677029 &1.06612 & 3.5537 & 4.21921\\
                                     & 5-loop & 0.275722 & 0.160396 & 1.19207 & 3.47991 &4.7682 \\
\hline
\multirow{2}{*}{$d=4$} & 4-loop & 0.841309 & 0.768325 & 1.99267 & 4.32652 & 4.98508\\
                                     & 5-loop & 0.858297 & 0.795556 & 2.02761 & 4.30235 & 5.255 \\
 \hline
\multirow{2}{*}{$d=5$} & 4-loop & 1.43379 & 1.41069 & 2.97067 & 5.14007 & 5.6558\\
                                     & 5-loop & 1.43578 & 1.41352 & 2.97374 & 5.13721 & 5.70711 \\
\end{tabular}
\caption{Scaling dimensions of operators in higher dimensions in the $(3,8)$ universality class, which follow from the GL description \eqref{eq:GL_action}. In the 4-loop and 5-loop results, we use the same Pad\'e approximants for any given operator as used in Sec.\ \ref{sec:5_loop}.
}
\label{tab38_MM_primaries_higher_d}
\end{table}

While $ \cO ^{(3)} _+$ is a Virasoro descendant in $d=2$, it cannot be a descendant in higher dimensions since we know that those are the operators with explicit derivatives acting on them. Therefore we expect it to be a new primary in $d>2$, and its dimension is shown in the table.

\section*{Acknowledgments}

We thank Shai Chester, Zohar Komargodski and Shu-Heng Shao for useful discussions. We are also grateful to the referee for very useful comments and for pointing out references.  
This work was supported in part by the US National Science Foundation under Grant No.~PHY-2209997 and by the Simons Foundation Grant No.~917464.

\appendix

\section{The $(3,10)$ minimal model} \label{sec:310_minimal_model}

In this appendix we provide more details on the $(3,10)$ minimal model. The operator dimensions and symmetries are presented in Table \ref{tab310_MM_primaries}. 

As mentioned before, the modular invariant relevant to us is the $D_6$ combination. In fact, it includes combinations of chiral and anti-chiral parts of the fields $\phi _{1,3} $, $\phi _{1,5} $, $\phi _{1,7} $, $\phi _{1,9} $, as well as the identity. Recall that $M(2,5)$ has only two primaries, the identity, as well as an operator $\phi ^{M(2,5)} $ having dimension $-\frac{2}{5} $. The operator $\phi _{1,5} $ clearly corresponds to $\phi ^{M(2,5)} $, while $\phi _{1,3} $ corresponds to two copies of this operator appearing in two copies of the $M(2,5)$ theory. The operator $\phi _{1,9} $ is the additional energy-momentum tensor that is present in the two copies of the theory.

Lastly, recall that \cite{Fei:2014xta} the flow from $M(3,10)$ to $M(3,8)$ is generated by the operator $\phi _1\phi _2^2+\phi _2\phi _1^2$. Using the equations of motion, this is the operator $\phi _1 \partial ^2\phi _2+\phi _2\partial ^2\phi _1$. Therefore, this operator has dimension $2+2\left(- \frac{2}{5}\right) =\frac{6}{5} $. The operator $\phi _{1,7} $ of $M(3,10)$ is precisely the operator generating the flow to $M(3,8)$.

To complete the description of $M(3,10)$, we collect the OPEs in this theory:
\begin{equation}
\begin{aligned}
& \phi _{1,2} \times \phi _{1,2} \sim 1+\phi _{1,3} ,\qquad && \phi _{1,2} \times \phi _{1,3} \sim \phi _{1,2} +i\phi _{1,4} ,\\
&\phi _{1,2} \times \phi _{1,4} \sim i\phi _{1,3} +\phi _{1,5} 
,\qquad && \phi _{1,2} \times \phi _{1,5} \sim \phi _{1,4} +\phi _{1,6} 
,\\
&\phi _{1,2} \times \phi _{1,6} \sim \phi _{1,5} +i\phi _{1,7} 
,\qquad && \phi _{1,2} \times \phi _{1,7} \sim i\phi _{1,6} +\phi _{1,8} 
,\\
&\phi _{1,2} \times \phi _{1,8} \sim \phi _{1,7} +\phi _{1,9} 
,\qquad && \phi _{1,2} \times \phi _{1,9} \sim \phi _{1,8} 
,\\
&\phi _{1,3} \times \phi _{1,3} \sim 1+\phi _{1,3} +i\phi _{1,5} 
,\qquad && \phi _{1,3} \times \phi _{1,4} \sim i\phi _{1,2} +\phi _{1,4} +\phi _{1,6} 
,\\
&\phi _{1,3} \times \phi _{1,5} \sim i\phi _{1,3} +\phi _{1,5} +i\phi _{1,7} 
,\qquad && \phi _{1,3} \times \phi _{1,6} \sim \phi _{1,4} +\phi _{1,6} +i\phi _{1,8} 
,\\
&\phi _{1,3} \times \phi _{1,7} \sim i\phi _{1,5} +\phi _{1,7} +\phi _{1,9} 
,\qquad && \phi _{1,3} \times \phi _{1,8} \sim i\phi _{1,6} +\phi _{1,8} 
,\\
&\phi _{1,3} \times \phi _{1,9} \sim \phi _{1,7} 
,\qquad && \phi _{1,4} \times \phi _{1,4} \sim 1+\phi _{1,3} +i\phi _{1,5} +\phi _{1,7} 
,\\
&\phi _{1,4} \times \phi _{1,5} \sim \phi _{1,2} +i\phi _{1,4} +i\phi _{1,6} +\phi _{1,8} 
,\qquad && \phi _{1,4} \times \phi _{1,6} \sim \phi _{1,3} +i\phi _{1,5} +\phi _{1,7} +\phi _{1,9} 
,\\
&\phi _{1,4} \times \phi _{1,7} \sim \phi _{1,4} +\phi _{1,6} +i\phi _{1,8} 
,\qquad && \phi _{1,4} \times \phi _{1,8} \sim \phi _{1,5} +i\phi _{1,7} 
,\\
&\phi _{1,4} \times \phi _{1,9} \sim \phi _{1,6}
,\qquad && \phi _{1,5} \times \phi _{1,5} \sim 1+\phi _{1,3} +i\phi _{1,5} +\phi _{1,7} +\phi _{1,9} 
,\\
&\phi _{1,5} \times \phi _{1,6} \sim \phi _{1,2} +i\phi _{1,4} +i\phi _{1,6} +\phi _{1,8} 
,\qquad && \phi _{1,5} \times \phi _{1,7} \sim i\phi _{1,3} +\phi _{1,5} +i\phi _{1,7} 
,\\
&\phi _{1,5} \times \phi _{1,8} \sim \phi _{1,4} +\phi _{1,6} 
,\qquad && \phi _{1,5} \times \phi _{1,9} \sim \phi _{1,5} 
,\\
&\phi _{1,6} \times \phi _{1,6} \sim 1+\phi _{1,3} +i\phi _{1,5}+\phi _{1,7} 
,\qquad && \phi _{1,6} \times \phi _{1,7} \sim i\phi _{1,2} +\phi _{1,4} +\phi _{1,6} 
,\\
&\phi _{1,6} \times \phi _{1,8} \sim i\phi _{1,3} +\phi _{1,5} 
,\qquad && \phi _{1,6} \times \phi _{1,9} \sim \phi _{1,4} 
,\\
&\phi _{1,7} \times \phi _{1,7} \sim 1+\phi _{1,3} +i\phi _{1,5} 
,\qquad && \phi _{1,7} \times \phi _{1,8} \sim \phi _{1,2} +i\phi _{1,4} 
,\\
&\phi _{1,7} \times \phi _{1,9} \sim \phi _{1,3} 
,\qquad && \phi _{1,8} \times \phi _{1,8} \sim 1+\phi _{1,3} 
,\\
&\phi _{1,8} \times \phi _{1,9} \sim \phi _{1,2} 
,\qquad && \phi _{1,9} \times \phi _{1,9} \sim 1
.
\end{aligned}
\end{equation}

\section{Some relations between (super-)Virasoro characters}\label{charproof}
In this appendix, we will prove the following two identities relating the Virasoro characters of $M(3,8)$ to a super-Virasoro character of $\SM(2,8)$:
\begin{align}\label{tobep}
\chi_{1,2}^V(\mathfrak q)-\chi_{1,6}^V(\mathfrak q)=1, \,\,\,\,\, \,\, \chi_{1,2}^V(\mathfrak q)+\chi_{1,6}^V(\mathfrak q)=\chi^R_{1, 4}(\mathfrak q) .
\end{align}
Here $\chi_{r,s}^V(\mathfrak q)$ denotes the Virasoro character corresponding to the primary field $\phi^{M(3,8)}_{r,s}$ in the $(3, 8)$ minimal model, given by \cite{rocha1985vacuum}
\begin{align}\label{Vchar}
\chi_{r,s}^V(\mathfrak q)=\prod_{n=1}^\infty \frac{1}{1-\mathfrak q^n}\sum_{m\in\mathbb Z}\sum_{\pm }\left(\pm\right) \mathfrak q^{\frac{1}{96} (48 m+8 r-2\mp 3 s) (48 m+8 r+2\mp 3 s)}
\end{align}
and $\chi^R_{r, s}(\mathfrak q)$ is the super-Virasoro character corresponding to the R sector superconformal primary field $\phi^{\SM(2,8)}_{r,s}$ in $\SM(2, 8)$, given by \cite{DiFrancesco:1988xz, Klebanov:2003wg}
\begin{align}\label{Rchar}
\chi_{r,s}^R(\mathfrak q)=
\prod_{n=1}^\infty \frac{1+\mathfrak q^n}{1-\mathfrak q^n}\sum_{m\in\mathbb Z}\sum_{\pm} \left(\pm\right)\mathfrak q^{\frac{1}{32} (16 m+4 r\mp s)^2}. \end{align}
Using eq. (\ref{Vchar}) for both $\phi^{M(3,8)}_{1,2}$ and $\phi^{M(3,8)}_{1,6}$, we obtain 
\small
\begin{align}\label{I11}
\chi_{1,2}^V-\chi_{1,6}^V&=\prod_{n=1}^\infty \frac{1}{1-\mathfrak q^n}\sum_{m\in\mathbb Z} \left(-\mathfrak q^{24 m^2-10 m+1}+\mathfrak q^{24 m^2+2m}-\mathfrak q^{24 m^2+14 m+2}+\mathfrak q^{24 m^2+26m+7}\right)\nonumber\\
&=\frac{\sum_{m\in\mathbb Z}- \mathfrak q^{\frac{3(-4m+1)^2-(-4m+1)}{2}}+\mathfrak q^{\frac{3(-4m)^2-(-4m)}{2}}-\mathfrak q^{\frac{3(-4m-1)^2-(-4m-1)}{2}}+\mathfrak q^{\frac{3(-4m-2)^2-(-4m-2)}{2}}}{\prod_{n=1}^\infty(1-\mathfrak q^n)}.
\end{align}
\normalsize
Notice that the sum on the R.H.S of eq.\ (\ref{I11}) can also be written as $\sum_{m\in\mathbb Z}(-)^m \mathfrak q^{\frac{3m^2-m}{2}}$, and the latter is exactly the series expansion of the infinite product $\prod_{n=1}^\infty(1-\mathfrak q^n)$, known as the pentagonal number theorem.  This yields the first identity in eq.\ (\ref{tobep}). For the sum of $\chi_{1,2}^V$ and $\chi_{1,6}^V$, we get similarly
\begin{align}
\chi_{1,2}^V+\chi_{1,6}^V=\frac{\sum_{m} \sigma_m\,  \mathfrak q^{\frac{3m^2-m}{2}}}{\prod_{n=1}^\infty(1-\mathfrak q^n)}
\end{align}
where $\sigma_m$ is equal to 1 when $m\equiv 0$ or 1 (mod 4), and $-1$ when $m\equiv 2$ or 3 (mod 4). It is straightforward to check that such a sign function can be realized as $\sigma_m=(-)^\frac{3m^2+m}{2}$, which further leads to 
\begin{align}\label{VV+}
\chi_{1,2}^V+\chi_{1,6}^V&=\frac{\sum_{m} (-)^m \mathfrak (-\mathfrak q)^{\frac{3m^2-m}{2}}}{\prod_{n=1}^\infty(1-\mathfrak q^n)}=\prod_{n\ge 1}\frac{1-(-\mathfrak q)^n}{1-\mathfrak q^n}\nonumber\\
&=\prod_{n\ge 1}\frac{\left(1-\mathfrak q^{2n}\right)\left(1+\mathfrak q^{2n-1}\right)}{1-\mathfrak q^n}=\prod_{n\ge 1}\left(1+\mathfrak q^{n}\right)\left(1+\mathfrak q^{2n-1}\right)
\end{align}
where we have used the pentagonal number theorem again. Next, we will show that the R.H.S of eq. (\ref{VV+}) is exactly the super-Virasoro character $\chi^R_{1,4}$:
\begin{align}
\chi^R_{1,4}=\prod_{n=1}^\infty \frac{1+\mathfrak q^n}{1-\mathfrak q^n}\sum_{m\in\mathbb Z} \left( \mathfrak q^{2 (2m)^2}-\mathfrak q^{2 (2m+1)^2}\right)=\prod_{n=1}^\infty \frac{1+\mathfrak q^n}{1-\mathfrak q^n}\sum_{m\in\mathbb Z} (-)^m {\mathfrak q}^{2\, m^2}. 
\end{align}
The infinite sum over $m$ yields the elliptic theta function $\theta_4 ({\mathfrak q}^2)$. Using its Jacobi triple product representation, we get  
\begin{align}
\chi^R_{1,4}&=\prod_{n=1}^\infty \frac{\left(1+\mathfrak q^n\right)\left(1-\mathfrak q^{4n}\right)\left(1-\mathfrak q^{4n-2}\right)^2}{1-\mathfrak q^n}\nonumber\\
&=\prod_{n=1}^\infty \frac{\left(1+\mathfrak q^n\right)\left(1-\mathfrak q^{4n-2}\right)}{1-\mathfrak q^n}\prod_{n=1}^\infty\left(1-\mathfrak q^{2(2n)}\right)\left(1-\mathfrak q^{2(2n-1)}\right)\nonumber\\
&=\prod_{n=1}^\infty \frac{\left(1+\mathfrak q^n\right)\left(1+\mathfrak q^{2n-1}\right)\left(1-\mathfrak q^{2n-1}\right)\left(1-\mathfrak q^{2n}\right)}{1-\mathfrak q^n}=\prod_{n\ge 1}\left(1+\mathfrak q^{n}\right)\left(1+\mathfrak q^{2n-1}\right).
\end{align}
This completes the proof of the second identity in eq.\ (\ref{tobep}).


\bibliographystyle{ssg}
\bibliography{GL_MM}

\begingroup\raggedright\begin{thebibliography}{10}

\bibitem{Belavin:1984vu}
A.~Belavin, A.~M. Polyakov, and A.~Zamolodchikov, ``{Infinite Conformal
  Symmetry in Two-Dimensional Quantum Field Theory},'' {\em Nucl.Phys.} {\bf
  B241} (1984) 333--380.

\bibitem{Friedan:1983xq}
D.~Friedan, Z.-a. Qiu, and S.~H. Shenker, ``{Conformal Invariance, Unitarity
  and Two-Dimensional Critical Exponents},'' {\em Phys. Rev. Lett.} {\bf 52}
  (1984) 1575--1578.

\bibitem{Zamolodchikov:1986db}
A.~B. Zamolodchikov, ``{Conformal Symmetry and Multicritical Points in
  Two-Dimensional Quantum Field Theory},'' {\em Sov. J. Nucl. Phys.} {\bf 44}
  (1986) 529--533.

\bibitem{Wilson:1971dc}
K.~G. Wilson and M.~E. Fisher, ``{Critical exponents in 3.99 dimensions},''
  {\em Phys.Rev.Lett.} {\bf 28} (1972) 240--243.

\bibitem{Hager:2002uq}
J.~S. Hager, ``{Six-loop renormalization group functions of O(n)-symmetric
  phi**6-theory and epsilon-expansions of tricritical exponents up to
  epsilon**3},'' {\em J. Phys. A} {\bf 35} (2002) 2703--2711.

\bibitem{Hsieh:2020uwb}
C.-T. Hsieh, Y.~Nakayama, and Y.~Tachikawa, ``{Fermionic minimal models},''
  {\em Phys. Rev. Lett.} {\bf 126} (2021), no.~19 195701,
  \href{https://arxiv.org/abs/2002.12283}{{\tt 2002.12283}}.

\bibitem{Fei:2016sgs}
L.~Fei, S.~Giombi, I.~R. Klebanov, and G.~Tarnopolsky, ``{Yukawa CFTs and
  Emergent Supersymmetry},'' {\em PTEP} {\bf 2016} (2016), no.~12 12C105,
  \href{https://arxiv.org/abs/1607.05316}{{\tt 1607.05316}}.

\bibitem{Grover:2013rc}
T.~Grover, D.~N. Sheng, and A.~Vishwanath, ``{Emergent Space-Time Supersymmetry
  at the Boundary of a Topological Phase},'' {\em Science} {\bf 344} (2014),
  no.~6181 280--283, \href{https://arxiv.org/abs/1301.7449}{{\tt 1301.7449}}.

\bibitem{Atanasov:2022bpi}
A.~Atanasov, A.~Hillman, D.~Poland, J.~Rong, and N.~Su, ``{Precision bootstrap
  for the $ \mathcal{N} $ = 1 super-Ising model},'' {\em JHEP} {\bf 08} (2022)
  136, \href{https://arxiv.org/abs/2201.02206}{{\tt 2201.02206}}.

\bibitem{Cardy:1985yy}
J.~L. Cardy, ``{Conformal Invariance and the Yang-lee Edge Singularity in
  Two-dimensions},'' {\em Phys.Rev.Lett.} {\bf 54} (1985) 1354--1356.

\bibitem{Fisher:1978pf}
M.~Fisher, ``{Yang-Lee Edge Singularity and $\phi^3$ Field Theory},'' {\em
  Phys.Rev.Lett.} {\bf 40} (1978) 1610--1613.

\bibitem{Kompaniets_2021}
M.~Kompaniets and A.~Pikelner, ``Critical exponents from five-loop scalar
  theory renormalization near six-dimensions,'' {\em Physics Letters B} {\bf
  817} (jun, 2021) 136331.

\bibitem{Borinsky:2021jdb}
M.~Borinsky, J.~A. Gracey, M.~V. Kompaniets, and O.~Schnetz, ``{Five-loop
  renormalization of ${\phi}^3$ theory with applications to the Lee-Yang edge
  singularity and percolation theory},'' {\em Phys. Rev. D} {\bf 103} (2021),
  no.~11 116024, \href{https://arxiv.org/abs/2103.16224}{{\tt 2103.16224}}.

\bibitem{Xu:2022mmw}
H.-L. Xu and A.~Zamolodchikov, ``{2D Ising Field Theory in a magnetic field:
  the Yang-Lee singularity},'' {\em JHEP} {\bf 08} (2022) 057,
  \href{https://arxiv.org/abs/2203.11262}{{\tt 2203.11262}}.

\bibitem{amslaurea11308}
N.~Amoruso, ``Renormalization group flows between non-unitary conformal
  models,'' Master's thesis.

\bibitem{Zambelli:2016cbw}
L.~Zambelli and O.~Zanusso, ``{Lee-Yang model from the functional
  renormalization group},'' {\em Phys. Rev. D} {\bf 95} (2017), no.~8 085001,
  \href{https://arxiv.org/abs/1612.08739}{{\tt 1612.08739}}.

\bibitem{Anninos:2021eit}
D.~Anninos and B.~M\"uhlmann, ``{The semiclassical gravitational path integral
  and random matrices (toward a microscopic picture of a dS$_{2}$ universe)},''
  {\em JHEP} {\bf 12} (2021) 206, \href{https://arxiv.org/abs/2111.05344}{{\tt
  2111.05344}}.

\bibitem{Lencses:2022ira}
M.~Lencs\'es, A.~Miscioscia, G.~Mussardo, and G.~Tak\'acs, ``{Multicriticality
  in Yang-Lee edge singularity},'' \href{https://arxiv.org/abs/2211.01123}{{\tt
  2211.01123}}.

\bibitem{Nakayama:2022svf}
Y.~Nakayama and K.~Kikuchi, ``{The fate of non-supersymmetric
  Gross-Neveu-Yukawa fixed point in two dimensions},''
  \href{https://arxiv.org/abs/2212.06342}{{\tt 2212.06342}}.

\bibitem{Fei:2014xta}
L.~Fei, S.~Giombi, I.~R. Klebanov, and G.~Tarnopolsky, ``{Three loop analysis
  of the critical O(N) models in $6-\epsilon$ dimensions},'' {\em Phys.Rev.}
  {\bf D91} (2015), no.~4 045011, \href{https://arxiv.org/abs/1411.1099}{{\tt
  1411.1099}}.

\bibitem{Fei:2014yja}
L.~Fei, S.~Giombi, and I.~R. Klebanov, ``{Critical $O(N)$ Models in
  $6-\epsilon$ Dimensions},'' {\em Phys.Rev.} {\bf D90} (2014) 025018,
  \href{https://arxiv.org/abs/1404.1094}{{\tt 1404.1094}}.

\bibitem{Fei:2015kta}
L.~Fei, S.~Giombi, I.~R. Klebanov, and G.~Tarnopolsky, ``{Critical Sp$(N)$
  models in $6 -{\epsilon}$ dimensions and higher spin dS/CFT},'' {\em JHEP}
  {\bf 09} (2015) 076, \href{https://arxiv.org/abs/1502.07271}{{\tt
  1502.07271}}.

\bibitem{Klebanov:2021sos}
I.~R. Klebanov, ``{Critical Field Theories with OSp$(1|2M)$ Symmetry},'' {\em
  Phys. Rev. Lett.} {\bf 128} (2022), no.~6 061601,
  \href{https://arxiv.org/abs/2111.12648}{{\tt 2111.12648}}.

\bibitem{Caracciolo:2004hz}
S.~Caracciolo, J.~L. Jacobsen, H.~Saleur, A.~D. Sokal, and A.~Sportiello,
  ``{Fermionic field theory for trees and forests},'' {\em Phys.Rev.Lett.} {\bf
  93} (2004) 080601, \href{https://arxiv.org/abs/cond-mat/0403271}{{\tt
  cond-mat/0403271}}.

\bibitem{2021CMaPh.381.1223B}
R.~{Bauerschmidt}, N.~{Crawford}, T.~{Helmuth}, and A.~{Swan}, ``{Random
  Spanning Forests and Hyperbolic Symmetry},'' {\em Communications in
  Mathematical Physics} {\bf 381} (Feb., 2021) 1223--1261,
  \href{https://arxiv.org/abs/1912.04854}{{\tt 1912.04854}}.

\bibitem{Narovlansky:2022ijq}
V.~Narovlansky, ``{Dualities between fermionic theories and the Potts model},''
  \href{https://arxiv.org/abs/2210.01847}{{\tt 2210.01847}}.

\bibitem{Castro-Alvaredo:2017udm}
O.~A. Castro-Alvaredo, B.~Doyon, and F.~Ravanini, ``{Irreversibility of the
  renormalization group flow in non-unitary quantum field theory},'' {\em J.
  Phys. A} {\bf 50} (2017), no.~42 424002,
  \href{https://arxiv.org/abs/1706.01871}{{\tt 1706.01871}}.

\bibitem{Kausch:1996vq}
H.~Kausch, G.~Takacs, and G.~Watts, ``{On the relation between Phi(1,2) and
  Phi(1,5) perturbed minimal models},'' {\em Nucl. Phys. B} {\bf 489} (1997)
  557--579, \href{https://arxiv.org/abs/hep-th/9605104}{{\tt hep-th/9605104}}.

\bibitem{Quella:2006de}
T.~Quella, I.~Runkel, and G.~M. Watts, ``{Reflection and transmission for
  conformal defects},'' {\em JHEP} {\bf 0704} (2007) 095,
  \href{https://arxiv.org/abs/hep-th/0611296}{{\tt hep-th/0611296}}.

\bibitem{2011NJPh...13d5006A}
E.~{Ardonne}, J.~{Gukelberger}, A.~W.~W. {Ludwig}, S.~{Trebst}, and
  M.~{Troyer}, ``{Microscopic models of interacting Yang-Lee anyons},'' {\em
  New Journal of Physics} {\bf 13} (Apr., 2011) 045006,
  \href{https://arxiv.org/abs/1012.1080}{{\tt 1012.1080}}.

\bibitem{Melzer:1994qp}
E.~Melzer, ``{Supersymmetric analogs of the Gordon-Andrews identities, and
  related TBA systems},'' \href{https://arxiv.org/abs/hep-th/9412154}{{\tt
  hep-th/9412154}}.

\bibitem{Nakayama:2021zcr}
Y.~Nakayama, ``{Is there supersymmetric Lee\textendash{}Yang fixed point in
  three dimensions?},'' {\em Int. J. Mod. Phys. A} {\bf 36} (2021), no.~24
  2150176, \href{https://arxiv.org/abs/2104.13570}{{\tt 2104.13570}}.

\bibitem{Gracey:2015tta}
J.~A. Gracey, ``{Four loop renormalization of $\phi^3$ theory in six
  dimensions},'' {\em Phys. Rev. D} {\bf 92} (2015), no.~2 025012,
  \href{https://arxiv.org/abs/1506.03357}{{\tt 1506.03357}}.

\bibitem{Dotsenko:1984nm}
V.~S. Dotsenko and V.~A. Fateev, ``{Conformal Algebra and Multipoint
  Correlation Functions in Two-Dimensional Statistical Models},'' {\em Nucl.
  Phys. B} {\bf 240} (1984) 312.

\bibitem{Dotsenko:1984ad}
V.~S. Dotsenko and V.~A. Fateev, ``{Four Point Correlation Functions and the
  Operator Algebra in the Two-Dimensional Conformal Invariant Theories with the
  Central Charge c \ensuremath{<} 1},'' {\em Nucl. Phys. B} {\bf 251} (1985)
  691--734.

\bibitem{Dotsenko:1985hi}
V.~S. Dotsenko and V.~A. Fateev, ``{Operator Algebra of Two-Dimensional
  Conformal Theories with Central Charge C \ensuremath{<}= 1},'' {\em Phys.
  Lett. B} {\bf 154} (1985) 291--295.

\bibitem{Zamolodchikov:1986gt}
A.~B. Zamolodchikov, ``{Irreversibility of the Flux of the Renormalization
  Group in a 2D Field Theory},'' {\em JETP Lett.} {\bf 43} (1986) 730--732.

\bibitem{Cordova:2019wpi}
C.~C\'ordova, K.~Ohmori, S.-H. Shao, and F.~Yan, ``{Decorated $\mathbb{Z}_{2}$
  symmetry defects and their time-reversal anomalies},'' {\em Phys. Rev. D}
  {\bf 102} (2020), no.~4 045019, \href{https://arxiv.org/abs/1910.14046}{{\tt
  1910.14046}}.

\bibitem{LeClair:1997gv}
A.~LeClair, A.~Ludwig, and G.~Mussardo, ``{Integrability of coupled conformal
  field theories},'' {\em Nucl. Phys. B} {\bf 512} (1998) 523--542,
  \href{https://arxiv.org/abs/hep-th/9707159}{{\tt hep-th/9707159}}.

\bibitem{Delfino:1997ya}
G.~Delfino and G.~Mussardo, ``{Nonintegrable aspects of the multifrequency
  Sine-Gordon model},'' {\em Nucl. Phys. B} {\bf 516} (1998) 675--703,
  \href{https://arxiv.org/abs/hep-th/9709028}{{\tt hep-th/9709028}}.

\bibitem{Calabrese:2001bm}
P.~Calabrese and A.~Celi, ``{Critical behavior of the two-dimensional N
  component Landau-Ginzburg Hamiltonian with cubic anisotropy},'' {\em Phys.
  Rev. B} {\bf 66} (2002) 184410,
  \href{https://arxiv.org/abs/cond-mat/0111118}{{\tt cond-mat/0111118}}.

\bibitem{Schoutens:1990vb}
K.~Schoutens, ``{Supersymmetry and Factorizable Scattering},'' {\em Nucl. Phys.
  B} {\bf 344} (1990) 665--695.

\bibitem{Ahn:1990uq}
C.-r. Ahn, ``{Complete S matrices of supersymmetric Sine-Gordon theory and
  perturbed superconformal minimal model},'' {\em Nucl. Phys. B} {\bf 354}
  (1991) 57--84.

\bibitem{Ahn:1993qa}
C.-r. Ahn, ``{Thermodynamics and form-factors of supersymmetric integrable
  field theories},'' {\em Nucl. Phys. B} {\bf 422} (1994) 449--475,
  \href{https://arxiv.org/abs/hep-th/9306146}{{\tt hep-th/9306146}}.

\bibitem{Moriconi:1995aj}
M.~Moriconi and K.~Schoutens, ``{Thermodynamic Bethe ansatz for N=1
  supersymmetric theories},'' {\em Nucl. Phys. B} {\bf 464} (1996) 472--491,
  \href{https://arxiv.org/abs/hep-th/9511008}{{\tt hep-th/9511008}}.

\bibitem{Ahn:2000tj}
C.-r. Ahn and R.~I. Nepomechie, ``{The Scaling supersymmetric Yang-Lee model
  with boundary},'' {\em Nucl. Phys. B} {\bf 594} (2001) 660--684,
  \href{https://arxiv.org/abs/hep-th/0009250}{{\tt hep-th/0009250}}.

\bibitem{Eichenherr:1985cx}
H.~Eichenherr, ``{Minimal Operator Algebras in Superconformal Quantum Field
  Theory},'' {\em Phys. Lett. B} {\bf 151} (1985) 26--30.

\bibitem{Bershadsky:1985dq}
M.~A. Bershadsky, V.~G. Knizhnik, and M.~G. Teitelman, ``{Superconformal
  Symmetry in Two-Dimensions},'' {\em Phys. Lett. B} {\bf 151} (1985) 31--36.

\bibitem{Friedan:1984rv}
D.~Friedan, Z.-a. Qiu, and S.~H. Shenker, ``{Superconformal Invariance in
  Two-Dimensions and the Tricritical Ising Model},'' {\em Phys. Lett. B} {\bf
  151} (1985) 37--43.

\bibitem{Cappelli:1986ed}
A.~Cappelli, ``{Modular Invariant Partition Functions of Superconformal
  Theories},'' {\em Phys. Lett. B} {\bf 185} (1987) 82--88.

\bibitem{DiFrancesco:1988xz}
P.~Di~Francesco, H.~Saleur, and J.~B. Zuber, ``{Generalized Coulomb Gas
  Formalism for Two-dimensional Critical Models Based on SU(2) Coset
  Construction},'' {\em Nucl. Phys. B} {\bf 300} (1988) 393--432.

\bibitem{Klebanov:2003wg}
I.~R. Klebanov, J.~M. Maldacena, and N.~Seiberg, ``{Unitary and complex matrix
  models as 1-d type 0 strings},'' {\em Commun. Math. Phys.} {\bf 252} (2004)
  275--323, \href{https://arxiv.org/abs/hep-th/0309168}{{\tt hep-th/0309168}}.

\bibitem{Witten:1982df}
E.~Witten, ``{Constraints on Supersymmetry Breaking},'' {\em Nucl. Phys. B}
  {\bf 202} (1982) 253.

\bibitem{Osborn_2018}
H.~Osborn and A.~Stergiou, ``Seeking fixed points in multiple coupling scalar
  theories in the $\epsilon$ expansion,'' {\em Journal of High Energy Physics}
  {\bf 2018} (may, 2018).

\bibitem{Yurov:1989yu}
V.~P. Yurov and A.~B. Zamolodchikov, ``{Truncated Conformal Space Approach to
  Scaling Lee-Yang Model},'' {\em Int. J. Mod. Phys. A} {\bf 5} (1990)
  3221--3246.

\bibitem{PhysRevB.102.014426}
F.~B. Ramos, M.~Lencs\'es, J.~C. Xavier, and R.~G. Pereira, ``Confinement and
  bound states of bound states in a transverse-field two-leg Ising ladder,''
  {\em Phys. Rev. B} {\bf 102} (Jul, 2020) 014426.

\bibitem{rocha1985vacuum}
A.~Rocha-Caridi, ``Vacuum vector representations of the Virasoro algebra,'' in
  {\em Vertex operators in mathematics and physics}, pp.~451--473.
\newblock Springer, 1985.

\end{thebibliography}\endgroup

\end{document}